\documentclass[a4paper,fleqn]{cas-sc}

\usepackage[authoryear,longnamesfirst]{natbib}

\def\tsc#1{\csdef{#1}{\textsc{\lowercase{#1}}\xspace}}
\tsc{WGM}
\tsc{QE}

\begin{document}
\let\WriteBookmarks\relax
\def\floatpagepagefraction{1}
\def\textpagefraction{.001}

\shorttitle{Screening Biomarkers for Biomedical Diagnosis}    

\shortauthors{Huang, Chen, Phoa, Lin, and Lin}  

\title [mode = title]{An Efficient Approach for Identifying Important Biomarkers for Biomedical Diagnosis}  



%

\author[1,2]{Jing-Wen Huang} 

\author[1]{Yan-Hong Chen}

\author[1]{Frederick Kin Hing Phoa}
\cormark[1]

\ead{fredphoa@stat.sinica.edu.tw}

\author[3]{Yan-Han Lin}

\author[3,4,5]{Shau-Ping Lin}
\cormark[1]
\ead{shaupinglin@ntu.edu.tw}

\cortext[1]{Corresponding author}

\affiliation[1]{organization={Institute of Statistical Science, Academia Sinica},
            addressline={No.128, Academia Road, Section 2, Nankang}, 
            city={Taipei},
            postcode={11529}, 
            country={Taiwan}}
\affiliation[2]{organization={Institute of Statistics, National Tsing Hua University},
            country={Taiwan}}
\affiliation[3]{organization={Institute of Biotechnology, National Taiwan University},
            country={Taiwan}}
\affiliation[4]{organization={Center for Systems Biology, National Taiwan University},
            country={Taiwan}}
\affiliation[5]{organization={Agricultural Biotechnology Research Center, Academia Sinica},
            country={Taiwan}}
            














\begin{abstract}
In this paper, we explore the challenges associated with biomarker identification for diagnosis purpose in biomedical experiments, and propose a novel approach to handle the above challenging scenario via the generalization of the Dantzig selector. To improve the efficiency of the regularization method, we introduce a transformation from an inherent nonlinear programming due to its nonlinear link function into a linear programming framework. We illustrate the use of of our method on an experiment with binary response, showing superior performance on biomarker identification studies when compared to their conventional analysis. Our proposed method does not merely serve as a variable/biomarker selection tool, its ranking of variable importance provides valuable reference information for practitioners to reach informed decisions regarding the prioritization of factors for further investigations.
\end{abstract}




\maketitle

\section{Introduction}
In the modern era with improved computational and storage capabilities, large-scale experiemtns with large number of factors are common to be conducted in many fields of research and decision-making scenarios, like manufacturing industrial product design, market strategy selection , drug discovery, blood screening, and other processes \citep{DL06}. A widely used rule-of-thumb principle called effect sparsity \citep{BM86} suggests that only a limited subset of factors among all are truly significant to the response of the experiment, so the importance of conducting a thorough and efficient factor screening procedure prior to the experiment cannot be overstated. A good screening approach significantly impacts the quality and relevance of subsequent analyses and decisions concluded from the experiment.

Most traditional experiments usually obtain a continuous-valued response in each trial, and so does the setup of the analysis in most screening experiments. However, when the response is no longer continuous in nature, the challenges associated with factor screening become pronounced. Although there exists some specialized techniques to address discrete-valued responses, they often come at a substantial computational cost, making their practical implementation cumbersome ad resource-intensive. In response to the growing demand for effective and computationally efficient factor screening methods, it is essential to develop a simple and robust programming structure tailored in specific for the experiments with discrete-valued response variables. Such structure should minimize computational burden to facilitate the systematic screening of factors that helps the practitioners to identify a list of significant factors towwards the response. Among all existing approaches, the regularization method stands out as a robust choice when coupled with cross-validation for the tuning parameters. While regularization methods hold a significant promise for effective variable selection, they are not without their challenges. One notable drawback is the computational burden they impose, which can hinder their practical implementations, especially in scenarios involving high-dimensional explanatory variables and discrete-valued response variable. 

This paper aims at exploring the challenges and searching for potential solutions associated with factor screening for experiments with non-continuous-valued responses, highlighting the significace of developing streamlined approaches for practical use. In specific, the main contribution of this work is to propose an easily accessible, efficient and novel approach, based on linear programming, for the above challenging scenario. Our proposed linear programming method offers a concise and user-friendly alternative to traditional regularization techniques. Although it may not always claim the best method in the realm of factor screening in all scenarios, it possesses a comparable ability to identify and prioritize essential factors, which has a potential to provide additional valuable insights and further simplify the screening process. The rest of the paper is arranged as follows. Section 2 briefly introduces some preliminary results. Section 3 presents the main theory and algorithm of this work. Section 4 demonstrates our method via data analysis and numerical analysis. A summary is given in the last section. 

\section{Preliminary} \label{sec:Preliminary}
\subsection{The Generalized Linear Model}
We consider a regular $k$-parameter exponential family, i.e., $\theta=\left(\begin{array}{c}
	\theta_1\\
	\vdots\\
	\theta_k
\end{array}\right)$ with common regularity conditions, its probability mass/density function is 
$$f(y;\theta)=\frac{1}{c(\theta)}h(y)e^{\theta^Tt(y)}.$$
Let $E(t(y))=\mu_t(\theta)$ and its inverse $\theta=g(\mu_t(\theta))$. We have a linear relationship $\theta_1=X\beta_1, \cdots, \theta_k=X\beta_k$, where the columns of $X$ are some known functions for the explanatory variables. When $k=1$, it reduces to the classical generalized linear model. The form of probability mass/density function for 1-parameter exponential family reduces to 
$$f(y)=e^{\frac{y\theta-b(\theta)}{a(\phi)}+c(y,\phi)},$$
where $E(Y)=\mu$ and $\theta=g(\mu)=X\beta.$

For example, given a binary data, $E(Y)=\mu$ is the probability of outcome $1$ and $1-\mu$ is the probability of outcome $0$. The canonical link function is 
$$\theta=g(\mu)=logit(\mu)=\left(\begin{array}{c}
	\ln(\frac{\mu_1}{1-\mu_1})\\
	\vdots\\
	\ln(\frac{\mu_n}{1-\mu_n})
\end{array}\right).$$
Then, the prediction of the mean of $Y$ is 
$$\hat{Y}=\left(\begin{array}{c}
	\hat{Y_1}\\
	\vdots\\
	\hat{Y_n}
\end{array}\right)=\left(\begin{array}{c}
	\frac{e^{x_1^T\hat{\beta}}}{1+e^{x_1^T\hat{\beta}}}\\
	\vdots\\
	\frac{e^{x_n^T\hat{\beta}}}{1+e^{x_n^T\hat{\beta}}}\\
\end{array}\right)=g^{-1}(X\hat{\beta}).$$ 

\subsection{Regularization Methods}
The most well-known regularization methods include the ridge regression \citep{HK68}, LASSO \citep{T96}, and the Dantzig selector \citep{CT07}. At their cores, these regularization techniques search for a delicate balance between two essential components: the fitting loss function and the norm of the model coefficients. Under normality assumption on the response, the ridge regression achieves the balance between residual sum of squares and the $l_2$ norm of the coefficients, LASSO gets right balance between residual sum of squares and the $l_1$ norm of the coefficients, and the Dantzig selector balance both the absolute gradient of the residual sum of squares and the $l_1$ norm of coefficients. By introducing a penalty term associated to the coefficient norm, they encourage simpler models that are less prone to overfitting, especially for those with high-dimensional explanatory variables.

Among these regularization methods, LASSO and the Dantzig selector demonstrate a unique and powerful capability in variable selection. By driving certain coefficients to exact zero, they facilitate the identification of the most pertinent predictors within the model. While the ridge regression typically finds its application in scenarios with normally-distributed response variables, LASSO can be generalized to handle a broader range of settings, including generalized linear models. Similarly, the Dantzig selector can also be extended to accommodate response variables following the 1-parameter exponential family distribution. Note that the ridge regression and LASSO are considered as non-linear programming methods, whereas the Dantzig selector can be recast as linear programming under normal response. However, when a non-normally distributed response variable is considered, the Dantzig selector also transitions to a non-linear programming framework in general. The simplest case for the Dantzig selector to extend to 1-parameter exponential family was studied in \cite{JR08}. For exponential family of the form $f(y)=e^{\frac{y\theta-b(\theta)}{a(\phi)}+c(y,\phi)}$ with its canonical link function $\theta=g(E(Y))=X\beta$, and $E(Y)=g^{-1}(X\beta)$, the solution of maximizing likelihood method satisfies
$$X^T(Y-\hat{E(Y)})=0$$
This programming becomes nonlinear because of the nonlinear link function. Thus, it is unavoidable to emply a complicated programming for regularization methods when the response is generalized, even though these methods offer a balance between model complexit and predictive accuracy.

\section{Main Result: Our Extension from Dantzig Selector}\label{sec:Main Results}
Following \cite{JR08}, we first generalize the Dantzig selector to the $k$-parameter exponential family. 	For exponential family of the form $f(y;\theta)=\frac{1}{c(\theta)}h(y)e^{\theta^Tt(y)}$ with its canonical link function $\theta=g(\mu_{t(Y)})=\left(\begin{array}{c}
		\theta_1\\
		\vdots\\
		\theta_k
	\end{array}\right)=\left(\begin{array}{c}
	X\beta_1\\
	\vdots\\
	X\beta_k
\end{array}\right)$ and $E(t(Y))=\mu_{t(Y)}$, the solution of maximizing likelihood method satisfies
	$$X^T(t_j(Y)-\hat{\mu}_{t_j(Y)})=0,$$
	for $j=1,\cdots,k$.
The proof of this result is provided in the appendix. Applying the above result leads to a straightforward extension from one-parameter exponential family to regular k-parameter exponential family. That is, the objective of extended Dantzig selector is also to 
\begin{equation}
	\hbox{ minimize }  ||\beta||_1 \hbox{ subject to } ||X^T(Y-\hat{Y})||\leq\delta,
	\label{GD}
\end{equation}
just like the original Dantzig selector for normal response and \cite{JR08}.

The Dantzig selector can be recast as a linear programming problem in traditional linear model \citep{PPX09}, but it becomes a nonlinear programming problem in generalized linear model when the canonical link function is not linear. We only deal with the binary response and the logit link function in this paper as an illustration to tackle this non-linear problem. This method is readily applied to other exponential family in a similar way with further investigations on the derivatives of its link function.

First, we approximate the function $\frac{e^{x_i'\beta}}{1+e^{x_i'\beta}}$ by its Taylor expansion to the second term at $\beta=(\beta_0, 0_k^T)^T$, ignore the term that is much smaller than others, and obtain a new objective
\begin{equation}
\hbox{ minimize } ||\beta^*||_1 \hbox{ subject to } ||X^T(Y-X\beta^*))||_{\infty}\leq\delta,
\label{Taylor_Y_hat}
\end{equation}
where $\beta^*$ is defined by $\beta=(\beta_0,\beta^*).$
The reason for the expansion at $\beta=(\beta_0, 0_k^T)^T$ is by the effect sparsity principle, which suggests that most of the effects $\beta$ are equal to zero and only a small number of them dominate the systematic part of the model. Comparing to other possible points, $\beta=(\beta_0, 0_k^T)^T$ is more robust under various true models. Although it becomes a different objective function, our numerical studies support this expansion. We provide the details about why and when our approximation holds in the appendix.

To show our choice of expanding $0$ as initial is adequately good, we perform  the successive linear programming method \citep{GS61}, which is a sequential method with several linear programmings to approximate the nonlinear programming solution, based on our above result. In successive linear programming, we solve the linear programming obtained by Taylor expansion at the solution of previous linear programming, and the sequence of solutions can converge to the nonlinear programming solution under some regularity assumptions.

To show if the objective function in (\ref{Taylor_Y_hat}) is linear, we express it in an equivalent form as follows. 

\begin{align}\label{LP}
	\hbox{ minimize } \left(\begin{array}{c}
		1_k\\
		0_k
	\end{array}\right)^T\left(\begin{array}{c}
		u\\
		u+\beta^*
	\end{array}\right)
\end{align}
subject to	
\begin{align}\label{LPC}
	\left(\begin{array}{cc}
		\frac{1}{4}X^TX & -\frac{1}{4}X^TX \\
		-\frac{1}{4}X^TX & \frac{1}{4}X^TX \\
		2I_k & -I_k\\
		\multicolumn{2}{c}{I_{2k}}
	\end{array}\right)\left(\begin{array}{c}
		u\\
		u+\beta^*
	\end{array}\right)\geq\left(\begin{array}{c}
		-X^Ty-\delta 1_k\\
		X^Ty-\delta 1_k\\
		O_k\\
		O_{2k}
	\end{array}\right).
\end{align}
where $1_k$ is a vector of ones with length $k$, $0_k$ is a vector of zeros with length $k$, $I_k$ is an identity matrix with dimension $k\times k$, and $O_k$ is a matrix of zeros with dimension $k\times k$. The equivalent representation is similar to that in \cite{PPX09}.

After we obtain the form of linear programming, we need to select a suitable tuning parameter $\delta$. We establish the profile plot with cut points uniformly scattered on the interval $[0, \delta_0]$, where $\delta_0=\max |X^TY|$, and perform cross validation with the selected criterion depending on the users to find out the best $\delta$. We include a factor on specific $\delta$ if its eventual shrink-to-zero position is equal to or larger than this current $\delta$. We can also rank the importance of factors based on their shrink-to-zero orders. In specific, the last one to shrink to zero implies the most important one, and the second to the last takes the second place, and so on. 

\section{Data and Performance Analysis}\label{sec:Data Analysis}
\subsection{Dataset and Results from Previous Analysis}
We use the dataset, S3 table from \cite{CPZetal17}, to verify our method. The dataset consists of 150 Parkinson's disease patients and 100 healthy controls. The disease diagnosis (Parkinson's disease or not) serves as the response, and 39 lipid subclasses, which has total 520 lipid species, are the explanatory variables. In addition, they recorded the gender of all patients. They first identified important factors by p-values and q-values (false discovery rates) from the single variable logistic regression. After that, they used domain knowledge and historical data to support two models with lipid species in GlcCer and GM3 lipid subclasses by gender. In other words, they consider a model with main effects of gender and lipid species, and two-factor interactions between gender and lipid species. By splitting data randomly into 90\% training data and 10\% testing data, they use support vector machine to train and test the area under the receiver operating characteristic (AUC), and repeated for many times. They got the average AUC 0.742 for male and 0.644 for female. By the analysis of logistic regression and the bootstrapping support vector machine analysis, they concluded that GM3 is an important factor for differentiating Parkinson's disease patients and healthy controls.

There are four doubious settings in the analysis of \cite{CPZetal17}. First, since this experiment is supersaturated in nature, a model with two-factor interaction is not appropriate at the screening stage due to the lack of degree of freedom. Second, they do not use all GlcCer and GM3 subclasses in the classification analysis. Third, we suspect that in their initial selection of biomarkers based on the p-values and q-values that resulted in the discovery of GlcCer and GM3, they included all patients rather than the common statistical practice of splitting into training and testing data. Fourth, they do not identify GlcCer directly as an important factor in their logistic screening stage, even though their bootstrapping classification result is good.

\subsection{Our Data Analysis by Dantzig Selector}

In this section, we demonstrate the analysis of this dataset by our Dantzig selector. The details of our procedure is given in the appendix.

First, the Dantzig selector identifies 10 lipid subclasses ranked in the importance order as follows.
\[\text{TG}=\text{Cer}>\text{GlcCer}>\text{GalCer}=\text{GM3}=\text{PS}>\text{dhSulf}=\text{PE}>\text{DG}=L\text{PI.}\] 
Note that Cer and monohexylCer (GlcCer+GalCer), which were the lipid subclasses in the GCase pathway and were implicated for their importance in Parkinsonism \cite{MMHetal13}, were high ranked in Dantzig selector but were not identified as candidates based on the original analysis in \cite{CPZetal17}, where only GM3 and TG were identified in their analysis. The evidence of the GCase activity as a risk of Parkinson's disease is also supported in \cite{ADQetal14} and \cite{BS16}.

Next, we employ the Dantzig selector to identify 87 important lipid species. Among all 33 lipid species identified from the logistic significant tests in \cite{CPZetal17}, 22 of them are overlapped with our results. Note that the 20 lipid species listed in the Table 2 of \cite{CPZetal17} are suggested via expert knowledge, in which only 2 of them are identified in their logistic significant tests, but we identify 4 out of 10 GlcCer and 5 out of 10 GM3 lipid species in our analysis. 

Moreover, if the same classification analysis is performed using our identified lipid species, we have an average AUC about 0.709 for the model without gender, which is larger than their average AUC 0.664 by using the factors in the Table 2 under the same data splitting, or their AUC  values from two genders. In fact, the common practice to perform a classification analysis is to select variables by  the 90\% training data only instead of the whole data, then the resulting average AUC is 0.6 for our Dantzig selector and 0.569 for the p-value and q-value selection method used in \cite{CPZetal17} under the same data splitting.

\section{Summary}\label{sec:Summary}

In this study, we generalize the Dantzig selector and apply it to analyze the experiment with regular k-parameter exponential family responses. We improve the efficiency of the regularization method by introducing a transformation from an inherent nonlinear programming associated with the Dantzig selector into a linear programming framework. We provide a comprehensive exposition of its functionality when applied to binary response scenarios, noting that its applicability extends naturally to other response types through analogous procedures.

Our formulation of the binary Dantzig selector streamlines the computational structure within the regularization framework, yielding superior performance in the context of biomarker identification studies compared to conventional analysis techniques in genomic experiments. This method can be readily adapted to address high-dimensional screening challenges across a spectrum of applications.

It is important to emphasize that our proposed method does not merely serve as a variable selection tool. This ranking of variables provides valuable reference information, aiding researchers in making informed decisions regarding the prioritization of factors for further investigations. In scenarios where budget constraints limit the feasibility of conducting experiments on all selected factors, the derived ranking facilitates a strategic approach. Researchers can opt to focus their efforts on variables with higher importance rankings, thus it optimizes resource allocation and maximizes research impact.

One interesting finding in our identification of 10 lipid subclasses is the importance of GalCer, which is an isomer of GlcCer that is technically difficult to differentiate with the original LC/ESI/MS/MS analytical platform used in \cite{MMHetal13}, and therefore considered the two isomers as one confounding item as monohexosylceramides to conduct preliminary examination in PD patients. Although they hypothesize that their findings of importance is driven by the level changes from GlcCer rather than GalCer, they suggest future studies for confirmation. Our identification rank shows via a data-driven analysis that GlcCer is a more important biomarker than GalCer towards the diagnosis of PD patients. However, we also suggest to have a follow-up experiment to disentangle the confounding relations between GlcCer and GalCer, and to study in specific the importance of GalCer towards the diagnosis of PD patients.

\section{Supplementary materials}
\textbf{Data Availability. } The published lipidomic data is available in the original paper \cite{CPZetal17}.

\appendix
\section{Appendix}
\subsection{Assumptions behind our Approximation}
Our linear approximation is not only supported by numerical studies and the reason that 0 is not too far away from true $\beta$. 
We can formulate this reason in a theoretical way.
Note that the Taylor expansion of link function in $i$th equation $\text{link}(x_i^T\beta)$ is
\begin{align*}
	\frac{e^{\beta_0}}{1+e^{\beta_0}}+\frac{e^{\beta_0}}{1+e^{\beta_0}}{x_i^*}^T\beta^*-\frac{1}{2}p_i(1-p_i)(2p_i-1)\beta^*x_i^*{x_i^*}^T\beta^*
\end{align*}
where $\beta=(\beta_0, \beta^*)$, $x_i=(1, x_i^*)^T$, $p_i=\frac{e^{tx_i^T\beta}}{1+e^{tx_i^T\beta}}$, and $0<t<1$. 
If the term $\frac{1}{2}p_i(1-p_i)(2p_i-1)\beta^*x_i^*{x_i^*}^T\beta^*$ is small enough to ignore relative to $\frac{e^{\beta_0}}{1+e^{\beta_0}}{x_i^*}^T\beta^*$ for all $i$, then it can be well approximated by the constraint $\|X^T(Y-X\beta)\|<\delta$. 

For example, since the maximum of $|p_i(1-p_i)(2p_i-1)|$ is 0.1 no matter what value of $t$, $x_i$, and $\beta$ are. 
Let \[r({x_i^*}^T\beta^*)=\frac{\frac{1}{2}p_i(1-p_i)(2p_i-1){\beta^*}^Tx_i^*{x_i^*}^T\beta^*}{\frac{e^{\beta_0}}{1+e^{\beta_0}}{x_i^*}^T\beta^*}.\]
Then $|r(x)|<0.1$ if $\frac{e^{\beta_0}}{1+e^{\beta_0}}\geq 0.35$ and $0.302<\frac{e^{x_i^T\beta}}{1+e^{x_i^T\beta}}<0.698$ for all $i$, meaning that the probability of getting a response 1 for a mean condition of the population is 0.35 and the probabilities of getting a response 1 for all patients are between 0.302 and 0.698. 
The range of probability is a reasonable assumption because it is similar to the assumption such that logistic regression is stably estimable.

\begin{figure}[!htb]
	\centering
		\includegraphics[width=0.7\textwidth]
	{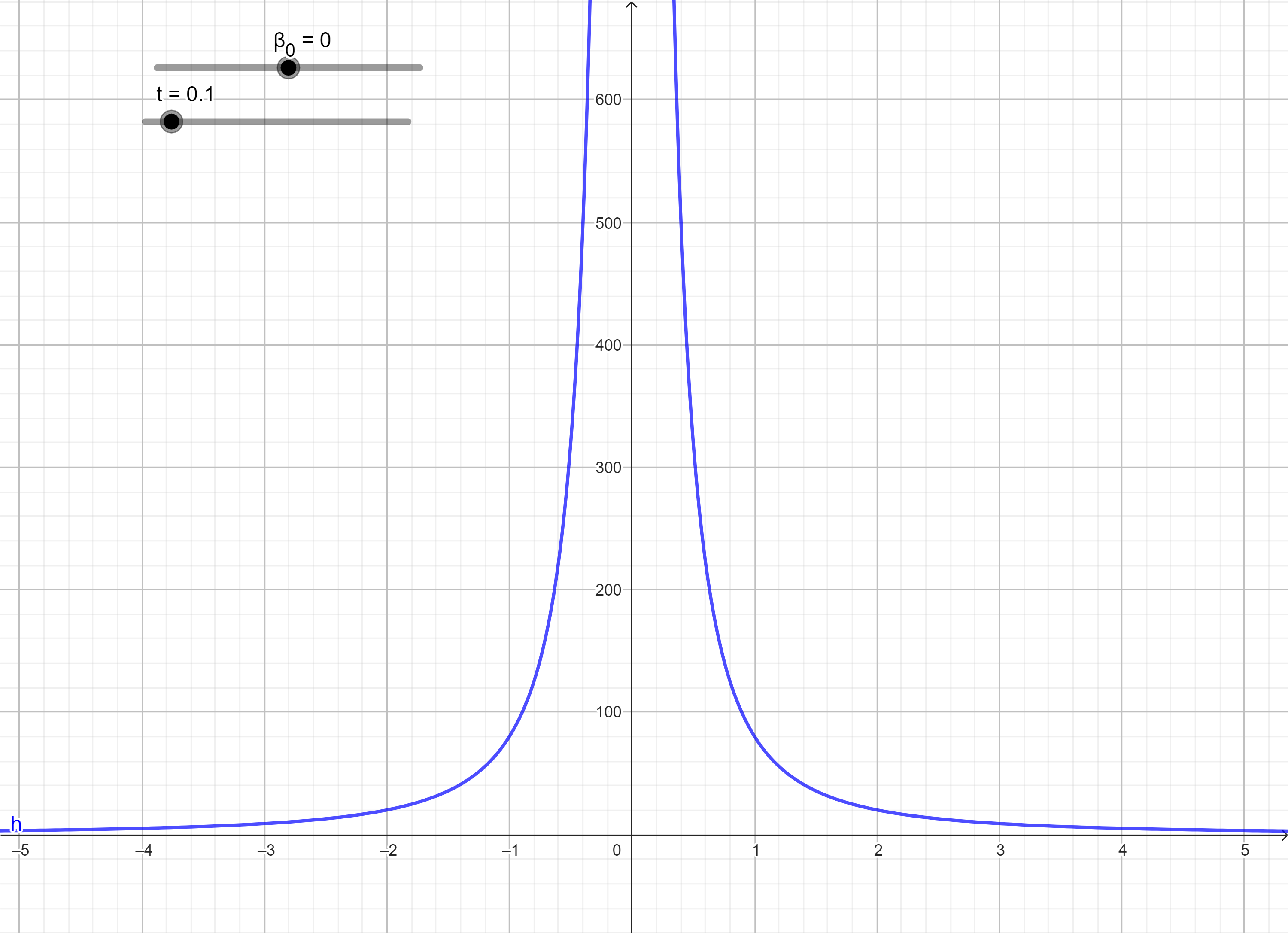}
	\caption{\label{fig:k_1_t_01} $\frac{1}{|r({x_i^*}^T\beta^*)|}$ when $\beta_0=0, t=0.1$} 
\end{figure}

\begin{figure}[!htb]
	\centering
	\includegraphics[width=0.7\textwidth]
	{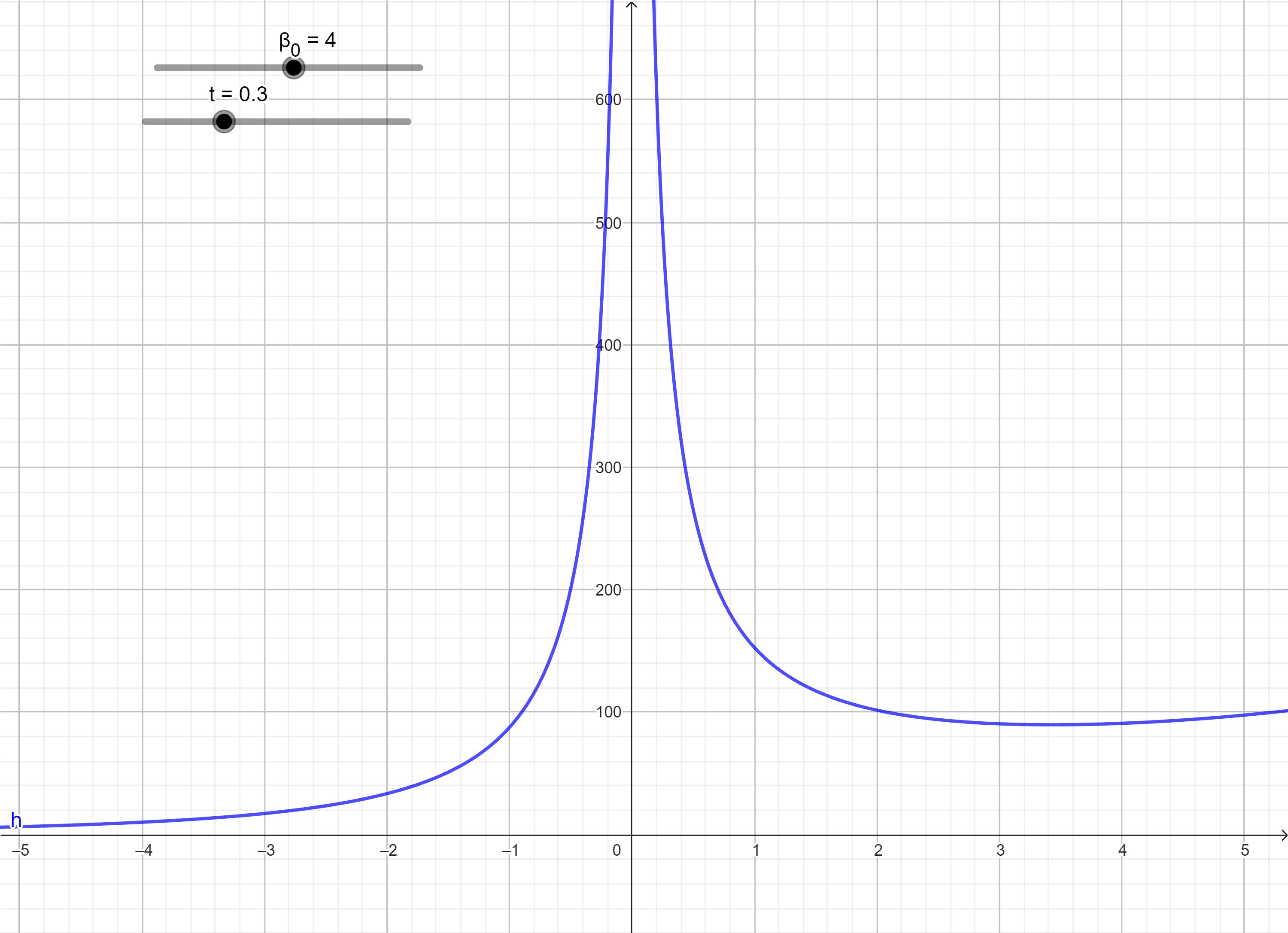}
	\caption{\label{fig:k_e4_t_03} $\frac{1}{|r({x_i^*}^T\beta^*)|}$ when $\beta_0=4, t=0.3$} 
\end{figure}

Actually, there exists a complex relation between $t$, $\beta_0$ and ${x_i^*}^T\beta^*$ such that the term $\frac{1}{2}p_i(1-p_i)(2p_i-1){\beta^*}^Tx_i^*{x_i^*}^T\beta^*$ is small enough to ignore relative to $\frac{e^{\beta_0}}{1+e^{\beta_0}}{x_i^*}^T\beta^*$ for all $i$. 
In Figure~\ref{fig:k_1_t_01}, we can further know if $0.29<\frac{e^{x_i^T\beta}}{1+e^{x_i^T\beta}}<0.71$, then $|r({x_i^*}^T\beta^*)|<0.01$. 
In Figure~\ref{fig:k_e4_t_03}, we can further know if $0.29<\frac{e^{x_i^T\beta}}{1+e^{x_i^T\beta}}<0.88$, , then $|r({x_i^*}^T\beta^*)|<0.01$. 
The range of probabilities depend on the value $\beta_0$ and $t$. 
Once you choose an $\epsilon$ you can accept for $|r({x_i^*}^T\beta^*)|<\epsilon$, there exists an assumption for $\beta_0$ such that $|r({x_i^*}^T\beta^*)|<\epsilon$. 
We can roughly judge whether the background knowledge of one dataset satisfies the corresponding assumption.


\subsection{A Proof of our Main Result}
Recall about our main result that for exponential family of the form $f(y;\theta)=\frac{1}{c(\theta)}h(y)e^{\theta^Tt(y)}$ with its canonical link function $\theta=g(\mu_{t(Y)})=(\theta_1, \dots, \theta_k)^T = (	X\beta_1, \dots, X\beta_k)^T$ and $E(t(Y))=\mu_{t(Y)}$, the solution of maximizing likelihood method satisfies $X^T(t_j(Y)-\hat{\mu}_{t_j(Y)})=0,$ for $j=1,\cdots,k$. To prove this result, we start with the joint likelihood 
$$L(\theta|y_1,\cdots,y_n)=\prod_{i=1}^n\frac{1}{c(\theta_i)}h(y_i)e^{\theta_i^Tt(y_i)}.$$
Then $\log L(\theta_1,\cdots,\theta_n|y_1,\cdots,y_n)=-\sum_{i=1}^n\log c(\theta_i) +\sum_{i=1}^n \log h(y_i)+\theta_i^T\sum_{i=1}^n t(y_i)$. We first claim that $\frac{\partial \log c(\theta_i)}{\partial \theta_j}=E(t_j)=\mu_{t_j}$. By definition, $c(\theta_j)=\int e^{\theta_j^Tt(y_i)}h(y_i)dy$. Hence, $\frac{\partial c(\theta_i)}{\partial \theta_j}=E(t_j)c(\theta_j)$, which implies $\frac{\partial \log c(\theta_i)}{\partial \theta_i}=\frac{\partial c(\theta_i)}{\partial \theta_i}\frac{1}{c(\theta_i)}= E(t(y_i)).$ \\
Finally, if $f(a)$ is a function from $\mathbb{R}^m$ to $\mathbb{R}$, and $a$ is a function of $b\in \mathbb{R}^n$, then $\frac{\partial f}{\partial b}=(\frac{\partial a}{\partial b})^T\frac{\partial f}{\partial a}$. We have
\begin{align*}
\frac{\partial \log L(\theta_1,\cdots,\theta_n|y_1,\cdots,y_n)}{\partial \beta_j}&=-\sum_{i=1}^n\frac{\partial \theta_i}{\partial \beta_j}^TE(t(y_i))+\sum_{i=1}^n\frac{\partial \theta_i}{\partial \beta_j}^T t(y_i)\\&=-X_j^T\mu_{t_j}(\theta)+X_j^Tt_j(y)=0,
\end{align*}
when we are solving maximum likelihood equation. 

\subsection{Details on Data Analysis}
\begin{figure}[!htb]
	\centering
	\includegraphics[width=0.7\textwidth]
	{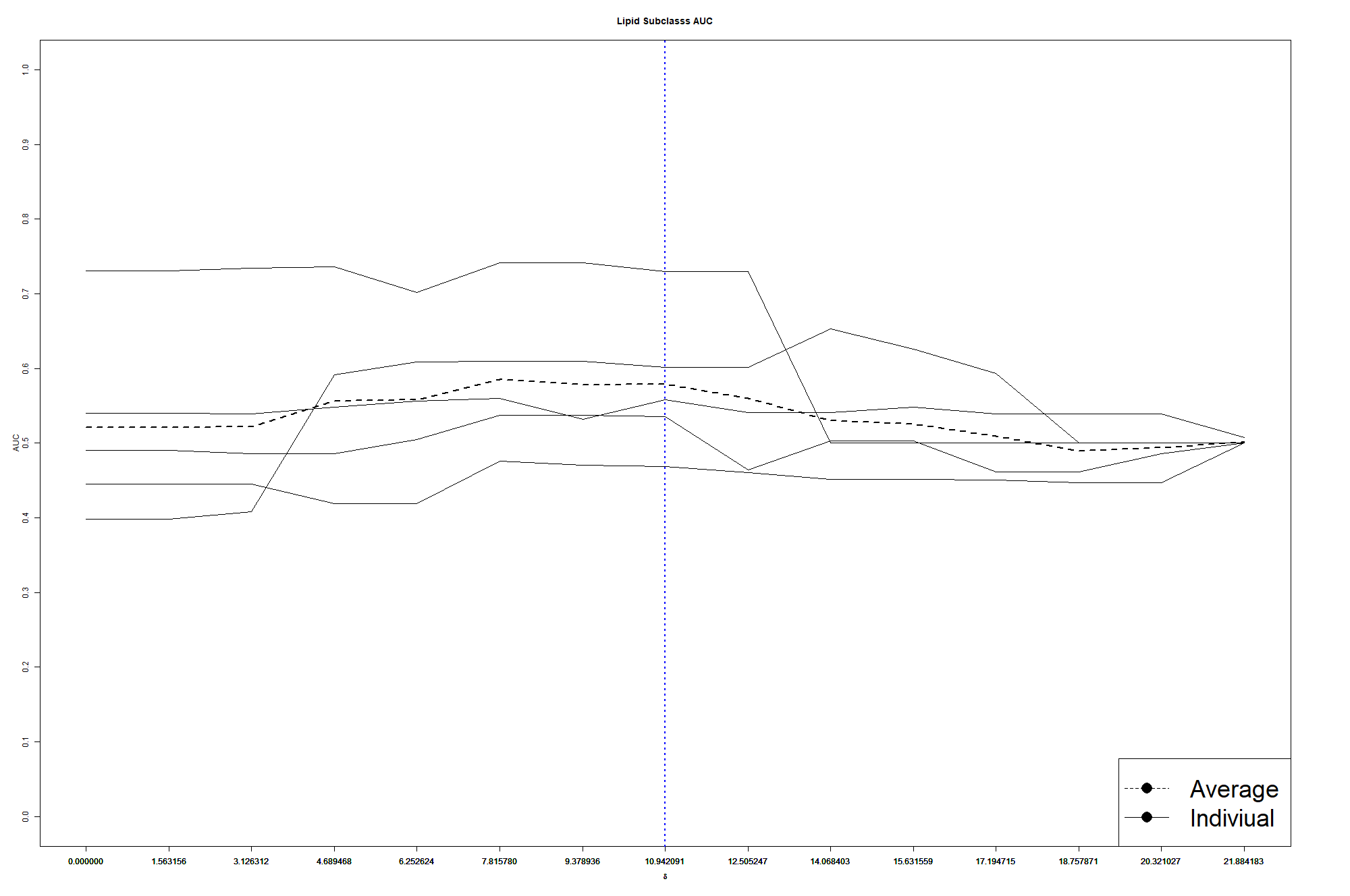}
	\caption{\label{fig:AUC_large} AUC from 5-fold Cross Validation} 
\end{figure}

\begin{figure}[!htb]
	\centering
	\includegraphics[width=0.7\textwidth]
	{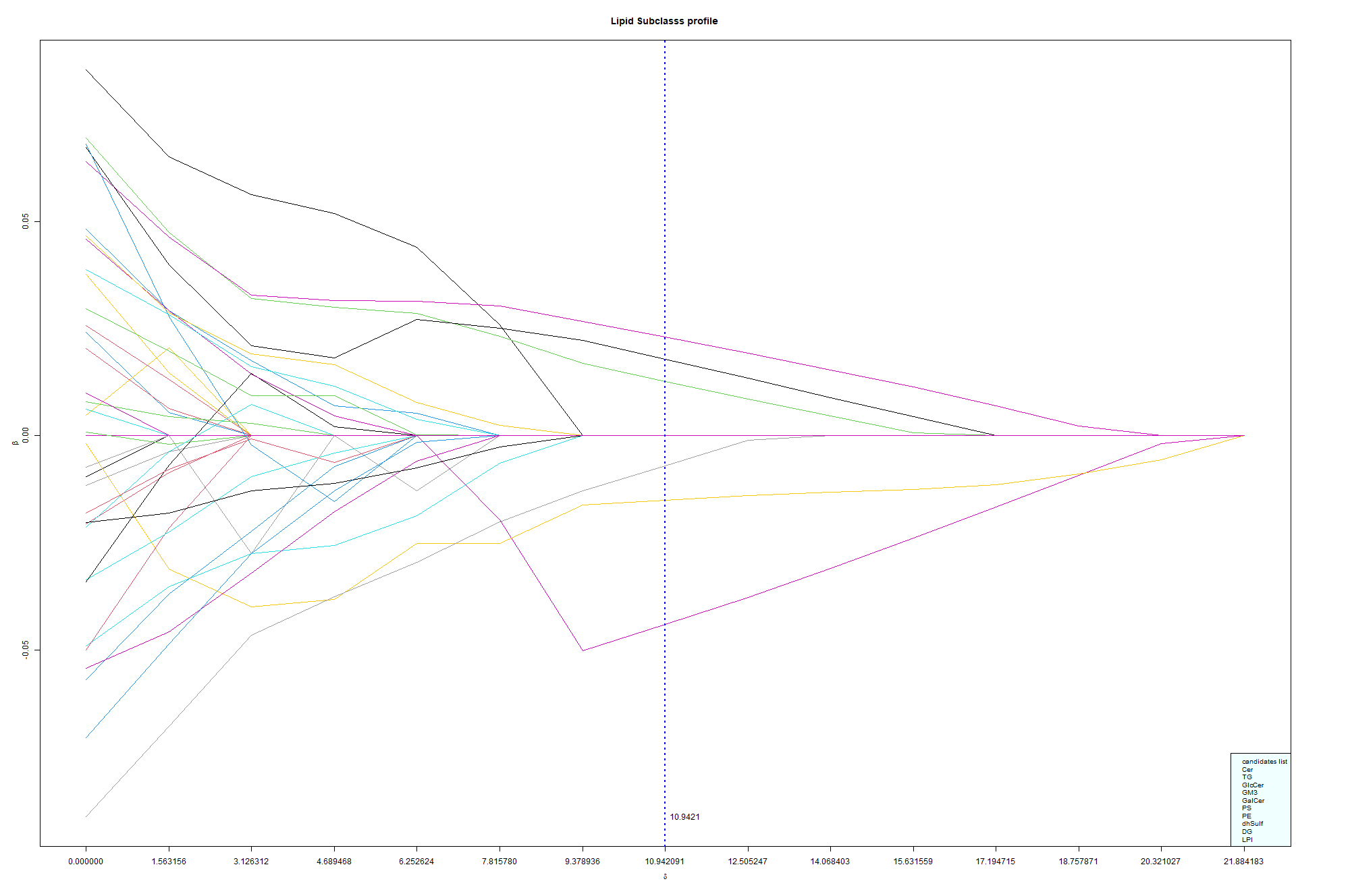}
	\caption{\label{fig:Lipid_large} Profile Plot for Lipid Subclasses} 
\end{figure}

We use the lipid subclasses data in \cite{CPZetal17} to illustrate our procedure. The procedure for analyzing lipid species data is the same. First, we calculate the AUC on uniformly scattered $\delta$ in $[0, \delta_0]$ by 5-fold cross validation like Figure~\ref{fig:AUC_large}. The x-axis is the value of $\delta\in[0,\delta_0]$, and the y-axis is the AUC value of 5-fold cross validations (solid lines) and their averages (dashed line). The blue vertical dashed line indicates the position of the best average AUC denoted as $\delta^*$. By a common practice in super saturated design analysis, we constrain the maximum number of selected factors to be smaller than $\frac{1}{4}$ number of the original factors, and the AUC comparison only takes on the values with this constraint.\\
After we find the position of $\delta^*$, we draw the profile plot to evaluate which factors shrink to zero on the right of this $\delta$ value. The complete profile plot is given in Figure~\ref{fig:Lipid_large}, but only the programming with $\delta\in[\delta^*, \delta_0]$ is needed in practice. In Figure~\ref{fig:Lipid_large}, the x-axis is the value of $\delta\in[0,\delta_0]$ and the y-axis is the solution of $\beta^*$ solved by the linear programming \eqref{LP} and its constraint \eqref{LPC}. The calculation suggests 10 lipid subclasses because their shrink-to-zero positions fall on the right of the blue dashed line $\delta^*$ in Figure~\ref{fig:Lipid_large}.

\printcredits

\bibliographystyle{cas-model2-names}

\bibliography{Main.bib}



\end{document}